\documentclass[letterpaper,10 pt,conference]{ieeeconf}  
\IEEEoverridecommandlockouts                              
\usepackage{graphics}
\usepackage{textcomp}
\usepackage{amsmath} 
\usepackage{mathtools}
\usepackage{amssymb}  
\usepackage{url}
\usepackage{commath}
\usepackage{caption}
\usepackage{array} 
\usepackage{color} 

\usepackage{footnote}
\makesavenoteenv{tabular}
\makesavenoteenv{table}
\makesavenoteenv{figure}

\usepackage{hyperref}
\usepackage{siunitx}
\usepackage{subcaption}
\usepackage{tikz}
\usepackage{booktabs}
\usepackage{multirow}
\usepackage{graphicx}
\usepackage{adjustbox}
\usepackage{comment}
\usepackage{dblfloatfix}

\title{\LARGE \bf
Investigating the Effect of Deictic Movements of a Multi-robot}

\author{Ahreum Lee, Wonse Jo, Shyam Sundar Kannan, and Byung-Cheol Min
\thanks{This research was supported in part by NSF CAREER Award IIS-1846221.}
\thanks{Ahreum Lee, Wonse Jo, Shyam Sundar Kannan, and Byung-Cheol Min are with SMART Lab, Department of Computer and Information Technology, Purdue University, West Lafayette, IN 47907, USA \tt\small{lahreum@purdue.edu, jow@purdue.edu, kannan9@purdue.edu, minb@purdue.edu}}%
}

\begin{document}
\maketitle
\begin{abstract}
Multi-robot systems are made up of a team of multiple robots, which provides the advantage of performing complex tasks with high efficiency, flexibility, and robustness. Although research on human-robot interaction is ongoing as robots become more readily available and easier to use, the study of interactions between a human and multiple robots represents a relatively new field of research. In particular, how multi-robots could be used for everyday users has not been extensively explored. Additionally, the impact of the characteristics of multiple robots on human perception and cognition in human multi-robot interaction should be further explored. In this paper, we specifically focus on the benefits of physical affordances generated by the movements of multi-robots, and investigate the effects of deictic movements of multi-robots on information retrieval by conducting a delayed free recall task. 

\end{abstract}


\section{Introduction}
Recent trends and advancements in robotics and automation have enhanced the presence of autonomous systems in human society and made robots part of our everyday lives. This has in turn increased interest in and focus on human-robot interaction. Robots are being used as social partners \cite{Logane20181511,ahmad2017adaptive}, educational assistants \cite{kanda2004interactive}, care-givers \cite{khosla2017human}, and more. 
These scenarios mostly involve interactions between a human and a single robot; however, while a single robot can handle a number of applications and issues, they are at times not so effective in handling complex scenarios. This has led to a growing role for multi-robots which are ``composed of large numbers of robots that can evolve in formation and adapt easily to multiple environments'' \cite{St-Onge:2019}.

In a multi-robot system, agents can share information, which can enhance the fault tolerance of the entire system. Also, a number of single robots can be easier to program and also cheaper to build compared to a single powerful robot that performs the same task \cite{yan2013survey}. Due to this flexibility and robustness to failure, multi-robots can perform laborious and dangerous tasks more efficiently than single robots, and have been applied in a broad range of applications such as reconnaissance, construction, environmental monitoring, exploration search, and infrastructure support \cite{kolling2016human}. 
Most studies to date have focused on functional aspects of multi-robots and the design and deployment of effective multi-robot systems, namely developing efficient methods to control multi-robot systems \cite{becker2013massive,setter2015haptic}.
However, only a few have considered how multi-robots could be used in human everyday life  \cite{ozgur2017cellulo,kim2017uur}. Moreover, there is less understanding of human perception and cognition in the context of multi-robot usage; the interactions between humans and multi-robots are less considered in multi-robot studies \cite{kolling2016human,kim2019swarmhaptics}.  

One potential area in which multi-robots could be embedded in everyday life is information display, taking advantage of the unique characteristics of multi-robots to present information in new ways \cite{suzuki2019shapebots,le2018dynamic}. In contrast with digital displays such as projectors and digital signage, a multi-robot being a group of physical objects can describe a 3D representation with its dynamic formation, including guiding communication by highlighting key information. To effectively use a multi-robot in such context, it is important to examine how the multi-robot affects the cognitive process and how people perceive the multi-robot. Several studies have examined how the movements of multi-robots affect emotion in the context of human-multi-robot interactions \cite{le2016zooids,santos2019motions}, but to the best of our knowledge, none have investigated the effects of multi-robots on cognition, particularly information recall. 

\begin{figure}[t]
  \centering
  \includegraphics[width=0.9\linewidth]{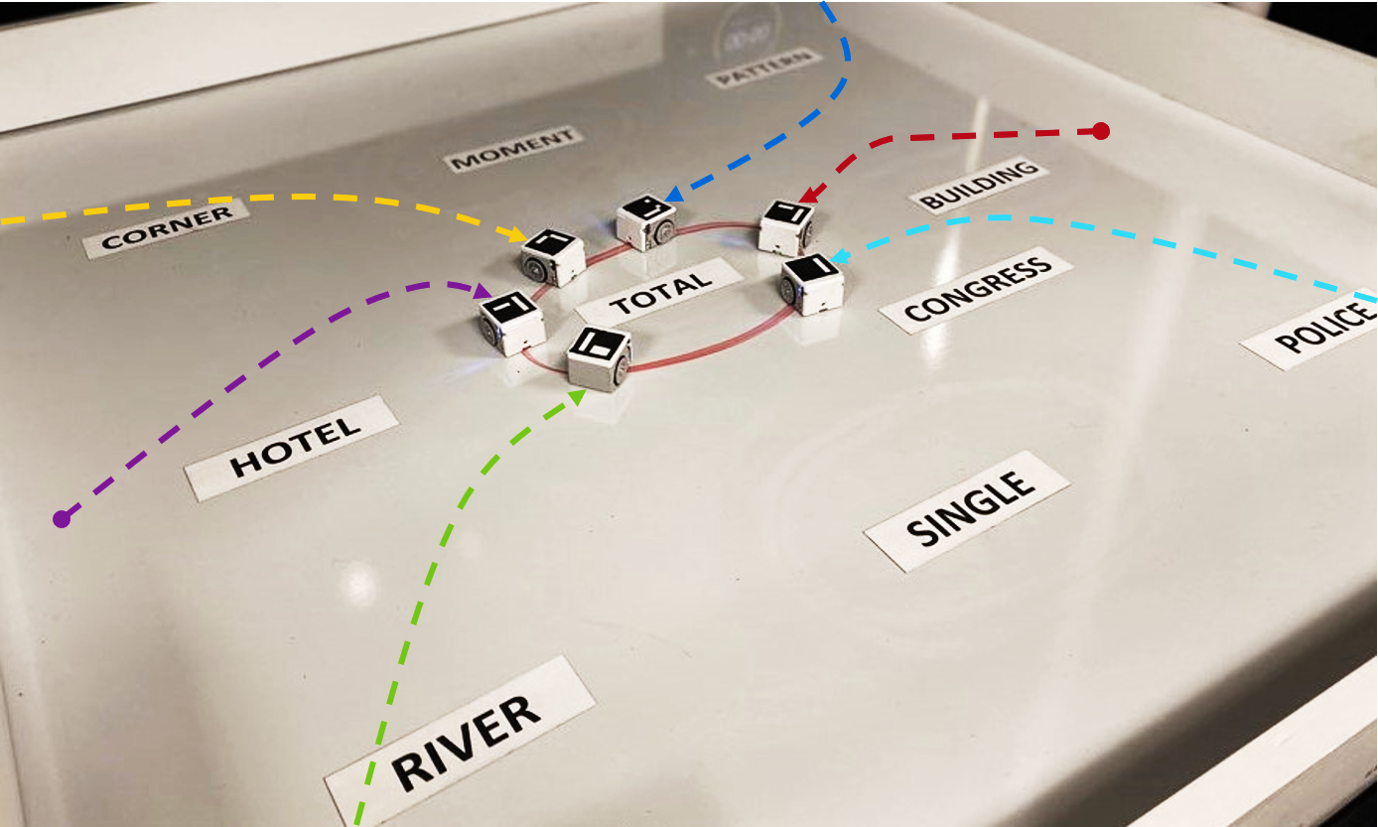}
  \caption{ A multi-robot based interface for deictic communication contexts.} \label{fig:concept}
\end{figure}

In this study, we conduct a user study with a multi-robot based interface (see Fig. \ref{fig:concept}). We particularly focus on the movement of a multi-robot as a deictic cue to highlight specific information and investigate the effects of the deictic cue created by a multi-robot on human cognition by examining memory performance with a delayed free recall task, which is a basic paradigm in the psychological study of memory. Specifically, we require participants to memorize 10 words, including one bonus word that is presented with a deictic movement of the multi-robot. 
The main contributions of this work are as follows: 1) developing a multi-robot-based interface for deictic communication, 2) designing experimental settings to investigate the effect of multi-robots on information retrieval using a delayed free recall task, and 3) conducting user studies on multi-robots with a mixed-methods approach. 

\section{Related works} 
\subsection{Multi-robot systems}

The study of multi-robot systems covers design, control, and implementation of a team of multiple robots \cite{parker1996multi,xi1993event,lucidarme2002implementation}; this research field started in the early 2000s and is still actively being studied. Especially from early 2010 to the present, the study of multi-robot systems has made rapid progress; it has become easier to create and control many robots due to the development of open source software and the Robot Operating System (ROS) \cite{quigley2009ros}, the miniaturization and dissemination of robot hardware \cite{sassone2016smart}, and the development of communication technologies \cite{wen2018swarm}. In addition, with the development of wireless communication \cite{wang2003ad}, control methods for a multi-robot through communication are being introduced. In fact, where research in the past focused on how to centralize multi-robot systems \cite{sanchez2002using}, in recent years the focus has shifted to achieving system scalability and stability through distributed control methods \cite{deng2019distributed}. 

As multi-robots become more diverse and the scope of their applicability increases, a number of studies have recently been conducted on human interaction with the multi-robots, called human multi-robot interaction. For example, \cite{podevijn2017effects}, \cite{naghsh2008analysis}, and \cite{setter2015haptic} all examined how multi-robots can execute commands and provide feedback to human operators. Moreover, several researchers have focused on the impacts of multi-robots on human psychology. \cite{podevijn2017effects} examined how the number of robots in a multi-robot system affected human psychophysiological status and \cite{santos2019motions} explored the effects of multi-robot movement and shape on human emotion. These studies provide positive recognition of multi-robot potential. However, the body of research on human multi-robot interaction is still very small compared to that for single robots. Two particular topic areas where research on human multi-robot interaction is lacking are in the context of human everyday life and the impact of multi-robots on human perception and cognition. With those gaps in mind, this paper, which examines human perception and cognition in the context of multi-robot usage, is very important and timely.

\subsection{Deictic cues}
In communication contexts, people use various non-verbal cues such as facial expressions, eye contact, and hand gestures to deliver their messages; these cues help convey one's emotion and emphasize what people want to communicate \cite{fillmore1982towards}. 
Moreover, a conversation that is based on verbal or visual information can be enriched by the inclusion of non-verbal gestures that provide additional context. Several studies have investigated how non-verbal cues affect information recall \cite{breazeal2005effects,so2012mnemonic}. One of the prototypical non-verbal cues is to refer an object by pointing out with a finger which provides a visual prominence on the object. The function of a deictic gesture is not just to orient another's attention towards a specific object but also to help them understand a relation between objects and contexts \cite{de2009towards,bernardos2016comparison}.

Similar strategies are also used to build visual prominence in information visualization, for example highlighting an object by enlarging its size, changing its color, or adding an outline  \cite{carenini2014highlighting,robinson2006highlighting}. High perceptual salience can also be built on essential information by employing a dynamic visual representation that temporally changes its appearance or position \cite{tversky2002animation,hoffler2007instructional}.  As simultaneous configuration of the properties of physical objects is limited, such dynamic representations have mostly been used to design virtual objects in digital contexts \cite{stusak2016exploring}. 

\begin{figure}
  \centering
  \includegraphics[width=1\linewidth]{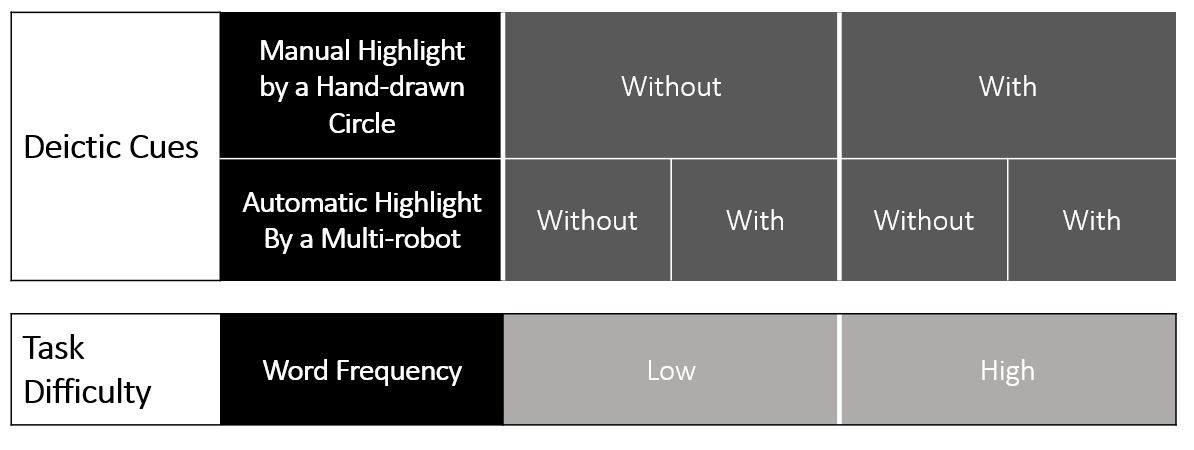}
  \caption{An experimental design with three within-subject factors; Two deictic cues: 1) a manual highlight by a hand-drawn circle and 2) an automatic highlight by a multi-robot, respectively, and task difficulty by controlling word frequency.} \label{fig:withinfactors}
\end{figure}

Attempts to design and develop a deictic cue with a dynamic representation have been extended to physical and tangible objects, which could incorporate physical affordances depicting their cognitive advantages \cite{klemmer2006bodies}. \cite{follmer2013inform} developed a dynamic shape display that can change its physical structure to guide a viewer's attention to a physical object.
Recently, the interest in developing a physical visualization medium have come to concentrate on the collective behaviors of multi-robot systems and several researchers have presented a new types of interfaces with a multi-robot which can effectively deliver information \cite{le2016zooids,suzuki2019shapebots,le2018dynamic}.
\cite{suzuki2019shapebots} designed shape-changing multi-robots that enhanced the flexibility and scalability of visualization; the authors focused on the hardware implementation and application scenarios. 
\cite{alonso2012image} developed an intuitive display that used multi-robots with controllable colors to visualize images and animations. To date, no investigation has been made into using the kinesthetic movements of multi-robots as deictic cues, nor into how the perception of information using a multi-robot compares against other deictic cues that have been well-used in communication and visualization contexts.
In this paper, we study how multi-robot systems can be used as a visual display capable of creating dynamic cues to emphasize textual information, how display using multi-robots affects human information recall in particular, and how humans perceive the deictic cues created by the multi-robot. 

\begin{figure*}
  \centering
  \includegraphics[width=0.9\linewidth]{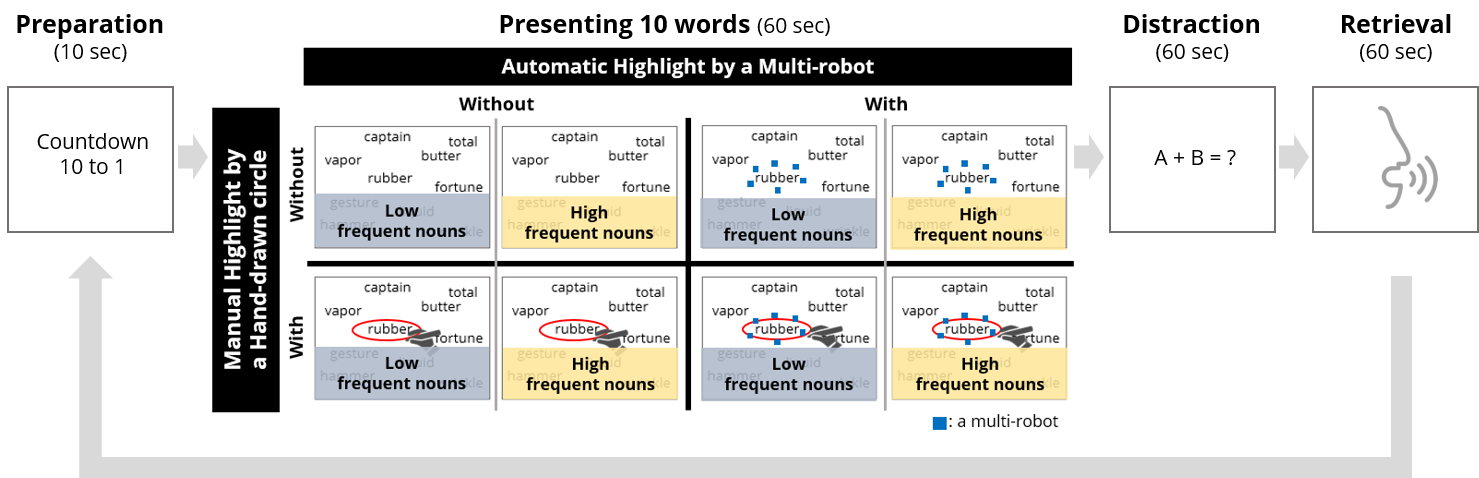}
  \caption{Experimental protocols.} 
  \label{fig:protocols}
\end{figure*}

\section{Our approach}  
With a delayed free recall task, we investigated the effect of deictic cues created by a multi-robot on information retrieval. A 2$\times$2$\times$2 within-subjects design was employed with three factors corresponding to a manual highlight (without and with a hand-drawn circle), an automatic highlight (without and with a multi-robot), and word frequency (low and high frequency) (see Fig. \ref{fig:withinfactors}). The first two factors were to create different types of deictic cues, and the last factor was to adjust task difficulty. All participants performed delayed free recall tasks in eight different conditions. 

\subsection{Free recall task}

In the typical delayed free recall tasks, each item to be memorized is presented alone for few seconds, and participants are asked to recall the items later without any hint \cite{murdock1974human,tulving1968theoretical}. Based on the classical paradigm of a delayed free recall task, we designed the delayed free recall tasks with four phases: preparation (10 seconds), encoding (60 seconds), distraction (60 seconds), and retrieval (60 seconds) (see Fig. \ref{fig:protocols}). However, we modified the typical delayed free recall task a little bit to investigate the effect of different types of deictic cues on information retrieval. 

First, we presented a list of items at the same time in the encoding phase. 
Specifically, for each task, a word set of 10 words was presented on designated locations of the whiteboard of the multi-robot interface. Among the 10 words, one word was randomly selected and denoted as a bonus word. 
Second, we asked participants to complete a dual-task: memorizing both a word set of 10 words and particularly a bonus word. In the retrieval phase, the participants were required to answer two questions: first, recall the word sets as fast as and as many as possible (``Can you recall the 10 words?''), and second, recall the bonus word only (``What was the bonus word?''). The participants were asked to answer the second question after they had responded to the first question. With the experimental settings, we tried to consider a trade-off effect on the task performances. Note a full demonstration of our modified delayed free recall tasks is available in the video\footnote{ YouTube link: \url{https://youtu.be/E0ETeqvvxPk}}. In this subsection, we further elucidate how we applied the three within-subjects factors to design the delayed free recall tasks.

\begin{figure}
    \centering
    \begin{subfigure}[b]{0.49\linewidth}
        \centering 
        \includegraphics[width=1\linewidth]{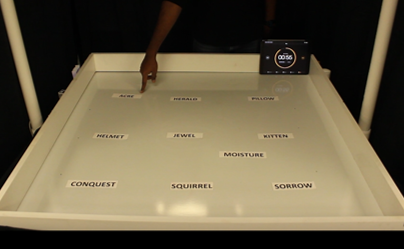} 
        \caption{Condition A}
        \label{}
    \end{subfigure}
    \begin{subfigure}[b]{0.49\linewidth}
        \centering 
        \includegraphics[width=1\linewidth]{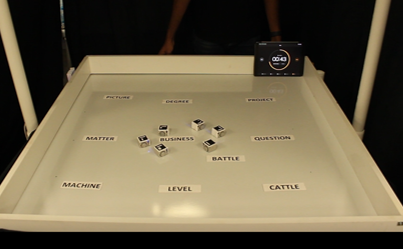} 
        \caption{Condition B}
        \label{}
    \end{subfigure}
    
    \begin{subfigure}[b]{0.49\linewidth}
        \centering 
        \includegraphics[width=1\linewidth]{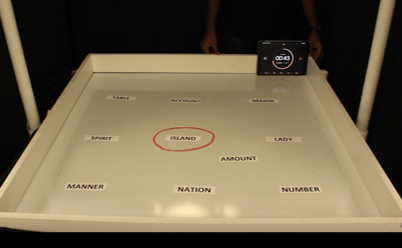} 
        \caption{Condition C}
        \label{}
    \end{subfigure}
    \begin{subfigure}[b]{0.49\linewidth}
        \centering 
        \includegraphics[width=1\linewidth]{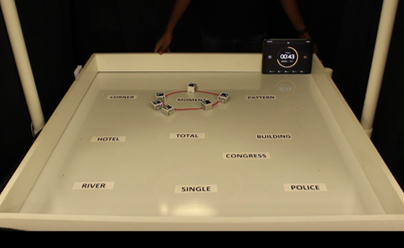} 
        \caption{Condition D}
        \label{}
    \end{subfigure}
  
    \caption{Four different conditions to define a bonus word; (a): a deictic gesture of the experimenter, (b): a multi-robot, (c): a hand-drawn circle, and  (d): both a multi-robot and a hand-drawn circle.} 
    \label{fig:4conditions}
\end{figure}

The way of presenting the bonus word was defined by the two within-subjects factors: 1) a manual highlight (i.e., with a hand-drawn circle), and 2) an automatic highlight (i.e., with a multi-robot). In this study, we used six commercial mobile robots to set a team of a multi-robot. To design a control condition of presenting a deictic cue with a hand-drawn circle and a multi-robot, we employed a common deictic gesture to highlight information, which is when a person points out the keyword with their index finger for few seconds \cite{neggers2001gaze}. 
As a result, four different conditions were defined by the way of highlighting a bonus word (see Fig. \ref{fig:4conditions}); \textit{condition A}: a deictic gesture of the experimenter (i.e., pointing out with their index finger for 5 seconds), \textit{condition B}: automatic highlight by a multi-robot, \textit{condition C}: manual highlight by a hand-drawn circle, and \textit{condition D}: highlight by both a multi-robot and a hand-drawn circle. 

\begin{figure*}
    \centering
    \begin{subfigure}[b]{0.2\linewidth}
        \centering 
        \includegraphics[width=1\linewidth]{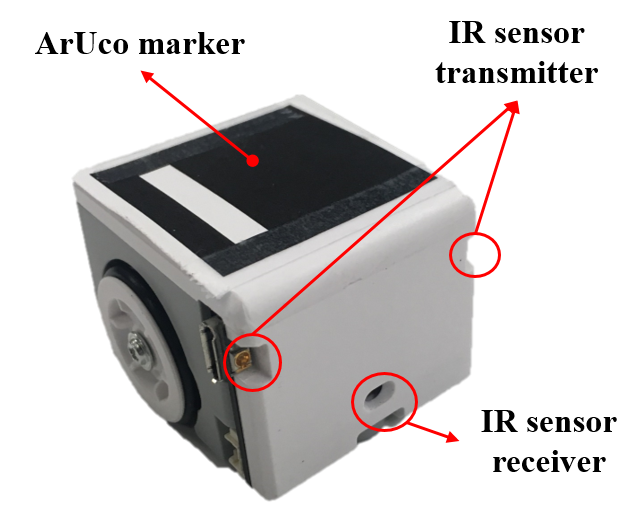} 
        \caption{Mobile robot platform}
        \label{fig:robot_platform}
    \end{subfigure}
    \begin{subfigure}[b]{0.35\linewidth}
        \centering 
        \includegraphics[width=1\linewidth]{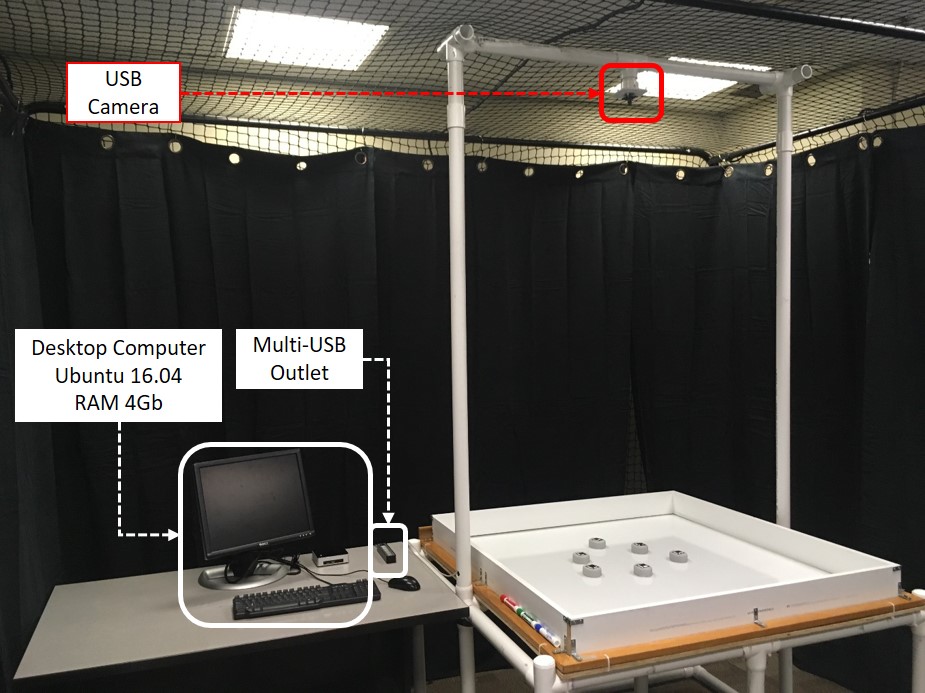} 
        \caption{Interactive testbed for multi-robots}
        \label{fig:mobile_testbed}
    \end{subfigure}
    \begin{subfigure}[b]{0.42\linewidth}
        \centering 
        \includegraphics[width=1\linewidth]{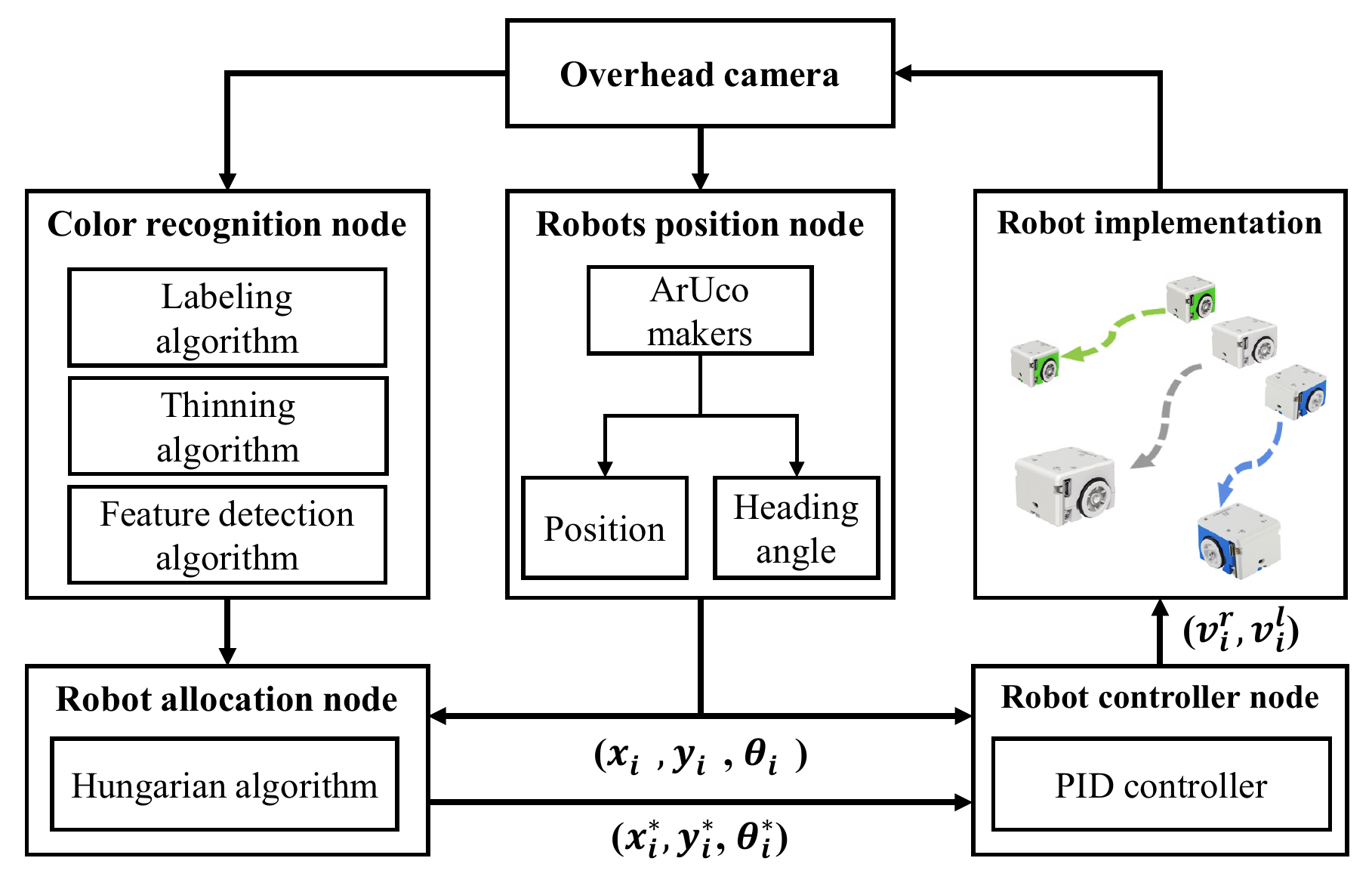} 
        \caption{ROS system diagram}
        \label{fig:system_diagram}
    \end{subfigure}
    
    \label{fig:whole_system_configuration}
    \caption{Details of the multi-robot interface used in this study.}

\end{figure*}

Due to different characteristics of the deictic cues, the exposure time of the bonus word had to be different over conditions. For the conditions with a hand-drawn circle (i.e., conditions C and D), the bonus word was pre-marked with the red circle, and participants could recognize and memorize the bonus word for a whole 60 seconds. For the control condition (i.e,. condition A), the bonus word was presented from the beginning but only for 5 seconds. Therefore, the participants had to keep memorizing what the bonus word is. 
Whereas, in the condition when the bonus word was only highlighted by a multi-robot (i.e., condition B), the participants were required to memorize the bonus word within 45 seconds. This is because it took about 15 seconds for the multi-robot to reach the bonus word from its initial position (we further elucidate how the multi-robot moves to the bonus word from its initial position in Section \ref{methods hwsw}). 
Therefore, the participants had a different amount of time to memorize the 10 words and bonus word depending on the conditions. In this regard, the two factors to highlight a bonus word with a deictic cue were not only different in their visual forms but also in their presentation modes. Although these four conditions did not generate the same situation for the participants to memorize the words, we focused on compare the effects of multi-robots to the deictic cues that people frequently use in daily lives.

The last factor was to adjust the task difficulty by presenting different frequency of words. As high-frequency words are easier to recall than low-frequency words \cite{hall1954learning,sumby1963word}, we defined the task difficulty in terms of the word frequency. With Kucera and Francis (K-F) frequency norms \cite{kuvcera1967computational}, we selected 40 low-frequency words (average and standard deviation of K-F frequency: 168.78, 83.13, respectively) and 40 high-frequency words (average and standard deviation of K-F frequency: 6.63, 3.45) from the Toronto Word Pool \cite{friendly1982toronto}. To set up word sets for eight tasks, we randomly set four groups of 10 words from the high-frequency words and another four groups of 10 words from the low-frequency words. 

\subsection{Design of a multi-robot interface} \label{methods hwsw}


We developed a multi-robot based interface where a multi-robot could present a deictic cue with its dynamic movement \cite{jo2019design}. In this section, we briefly describe how we designed the multi-robot interface and elucidate how we create deictic movements of a multi-robot in conditions B and D.
The hardware system consists of a commercial mobile robot platform (see Fig. \ref{fig:robot_platform}) and an interactive testbed (see Fig. \ref{fig:mobile_testbed}). 

\begin{figure*}
    \centering
    \begin{subfigure}[b]{0.195\linewidth}
        \centering 
        \includegraphics[width=1\linewidth]{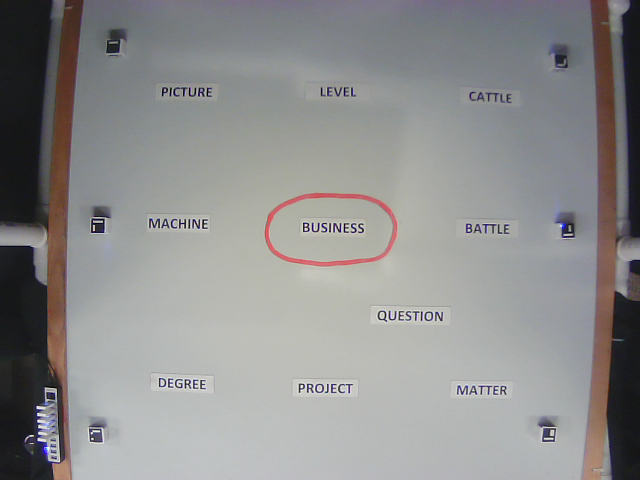} 
        \caption{}
        \label{fig:initial_pose}
    \end{subfigure}
    \begin{subfigure}[b]{0.195\linewidth}
        \centering 
        \includegraphics[width=1\linewidth]{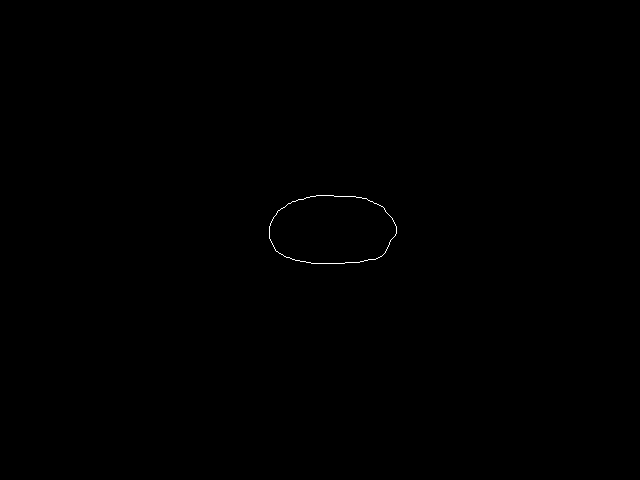} 
        \caption{}
        \label{fig:skeleton}
    \end{subfigure}
    \begin{subfigure}[b]{0.195\linewidth}
        \centering 
        \includegraphics[width=1\linewidth]{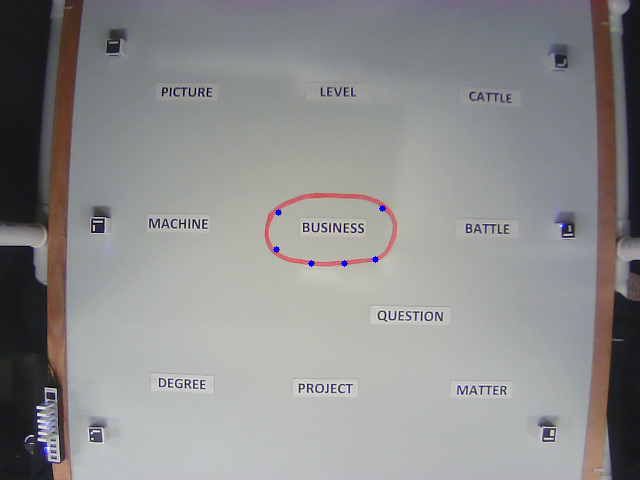} 
        \caption{}
        \label{fig:corner}
    \end{subfigure}
    \begin{subfigure}[b]{0.195\linewidth}
        \centering 
        \includegraphics[width=1\linewidth]{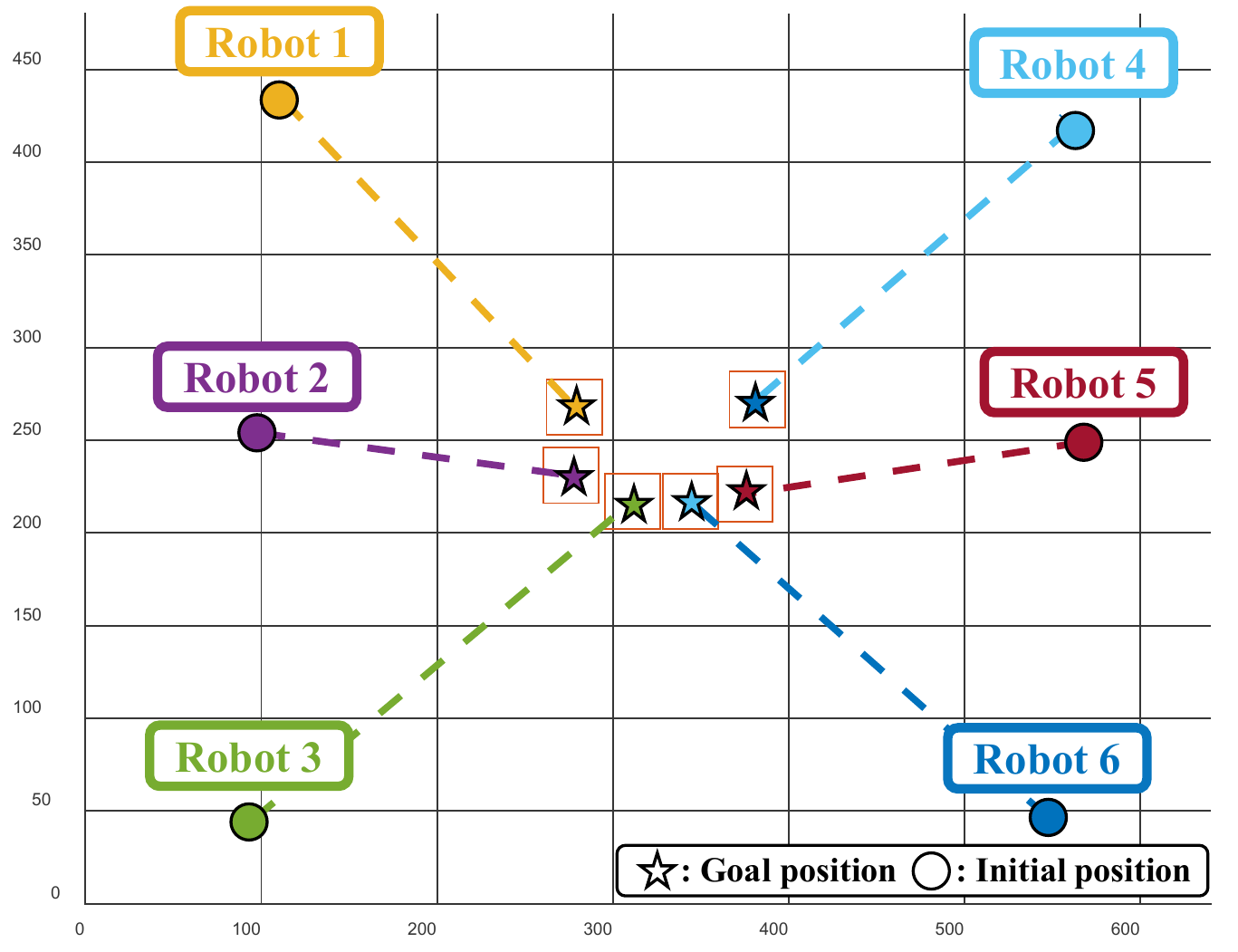} 
        \caption{}
        \label{fig:hungarian}
    \end{subfigure}
    \begin{subfigure}[b]{0.195\linewidth}
        \centering 
        \includegraphics[width=1\linewidth]{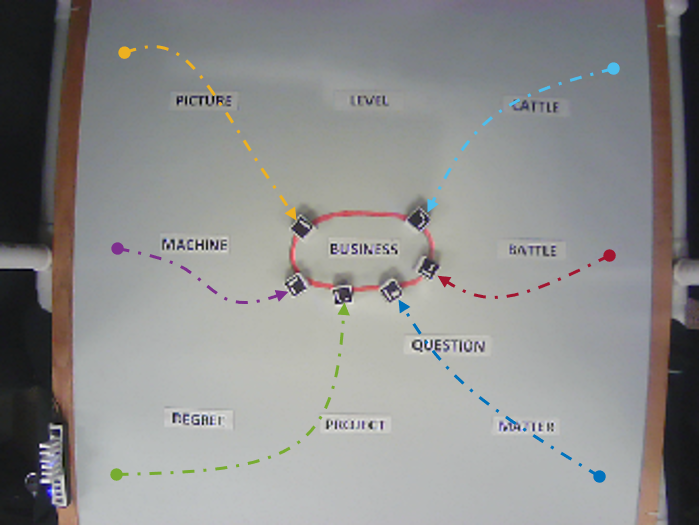} 
        \caption{}
        \label{fig:finial_pose}
    \end{subfigure}

    \caption{Example of the image-processing procedures; (a) the initial image showing the positions of the robots and the hand-drawn circle, (b) a thinning algorithm (skeletonization), (c) a corner detection algorithm, (d) robot allocation with Hungarian algorithm, and (e) results of the example study in case of the condition D.}
    \label{fig:image_processing}
\end{figure*}

The multi-robots used in this study consist of six commercial mobile robot platforms. The robot platform has differential wheels operated by DC (Direct Current) motors. The overall size of the robot is 35mm$\times$30mm$\times$40mm (width$\times$height$\times$depth) and its weight is 30g. The robots can detect objects using infrared (IR) sensors mounted on the platform and automatically avoid collisions and maintain relative distances from one another. The robots are controlled by the main computer system via Bluetooth communication \cite{kim2008roboid}.

For the interactive testbed platform, there is a square whiteboard where a multi-robot can move and people can present a word set which was printed on magnetic papers to attach on the board and draw a manual deictic cue (i.e., a hand-drawn circle) using a marker pen. The overall size of the testbed platform is 1.3m$\times$3m$\times$1.3m (width$\times$height$\times$length) which weighs about 50kg. Additionally, PVC pipes are used to build the testbed frame where an overhead camera (i.e., an USB camera) is mounted. The overhead camera consistently streams the global view of the whiteboard into the main computer, and then the streamed images are used for image processing techniques using open computer vision (OpenCV) libraries \cite{bradski2008learning} to track ArUco markers and recognize a hand-drawn circle on the whiteboard.
The overall system is processed in the main computer having an Intel Core i7 CPU and 4 GB RAM and utilizes the Robot Operating System (ROS). The ROS is one of the popular middlewares in the robotics field that is to enable various nodes (or programs) to share data using pre-defined protocols called \textit{standard\_message} with a global timer for data synchronization \cite{quigley2009ros}. In our system, we designed four nodes in the ROS system to enable multi-robots to move from their initial positions to the bonus word (i.e., condition B) or to the hand-drawn circle (i.e., condition D) as shown in Fig. \ref{fig:system_diagram}: robot position node, color recognition node, robot allocation nodes, and robot controller node. 

The robot position node was to continuously track robots' current position ($x_{i}, y_{i}$) and heading angle ($\theta_{i}$), $i\in{1,..,n}$ where $n=6$, through an ArUco marker on the top of each robot. The marker was to assign the robot's identifier and coordinate \cite{garrido2014automatic}. The initial positions of the robots were set to six positions where are 10cm away from the edge of the word set as shown in Fig. \ref{fig:initial_pose}.  

The color recognition node was applied to recognize the position of the hand-drawn circle for the condition D. As the first step, the image data obtained from the overhead camera were classified according to the color of the hand-drawn circle. Next, a thinning algorithm (or skeletonization) was applied on the targeted color label data to remove redundant pixels as depicted in Fig. \ref{fig:skeleton}. Then, a corner detection algorithm was applied to extract distinct points from the image data \cite{bradski2008learning} as shown in Fig. \ref{fig:corner}. The six distinct points became the goal positions for the robots in this study. However, the color recognition node was not necessary in the condition B as the bonus word was only presented by a multi-robot. Instead, we used predetermined six positions around the bonus word to define the goal positions of the multi-robot. 


The robot allocation node was to allocate each robot to the short-distance-based optimal goal position ($x^{*}_{i}$, $y^{*}_{i}$, $\theta^{*}_{i}$) calculated by the Hungarian algorithm \cite{kuhn1955hungarian}, where the goal positions were the six distinct points extracted from the color recognition node. As the initial positions of the multi-robot were dispersed (i.e., 10 cm away from the edge of the words) and the goal position was relatively located in the center of the board, the multi-robot moved from outside to inside of the whiteboard as illustrated in Fig. \ref{fig:hungarian}.

The robot controller node was to generate each wheel velocity ($v^{r}_{i}$, $v^{l}_{i}$) using a proportional–integral–derivative controller (PID) based on the kinematic model of the mobile robot, which is one of the traditional control methods in the robotics field \cite{mihelj2019mobile}.
The PID controller was designed to enable multi-robots to reach their goal positions by minimizing the position errors between their current positions ($x_{i}, y_{i}$, $\theta_{i}$) and the goal positions ($x^{*}_{i}$, $y^{*}_{i}$, $\theta^{*}_{i}$). As a result, the multi-robot can autonomously reach the pre-defined positions in the condition B or the allocated goal positions extracted from the robot allocation node in the condition D to highlight a bonus word on the whiteboard within about 15 seconds as shown in Fig. \ref{fig:finial_pose}.


 
\subsection{Participants}
40 participants (22 males and 18 females, average age: 25.03) were recruited through flyers and snowball sampling \cite{biernacki1981snowball}. All participants were either undergraduate or graduate students who were fluent in English. When we asked their previous experience with robots, 20 participants had experienced to interact with robots. They did not have any neurological disorders or any other conditions that may have affected their performance in the experiment. Each participant was compensated \$5 for participating in the study. This study was approved by the University’s Institutional Review Board (Purdue IRB Protocol: \#1902021821). 

\subsection{User Study Procedure}
Before engaging with the main experiment, participants received an overview of the study from an experimenter and signed the informed consent form. After signing their consent form, participants were asked to fill out a demographic questionnaire about gender, age, education level, self-reported English fluency, and previous experience with robots. Then, the participants were guided into the experiment space where the multi-robot interface was set up.
The experimenter further elaborated on the entire procedures constituting eight delayed free recall tasks. Each delayed free recall task was composed of 
four different phases: preparation (10 secs), encoding (i.e., presenting a word set) (60 secs), distraction (60 secs), and retrieval (60 secs) (see Fig. \ref{fig:protocols}). The materials for the preparation and distraction phases were presented on the screen which was on an opposite side of the multi-robot interface. Therefore, the participants had to turn around whenever they completed the preparation and encoding phases. This was purposely designed to strictly control the time for each phase. 

The participants were informed that the bonus word would be presented in four different ways in the encoding phase (see Fig. \ref{fig:4conditions}) and they would be asked to answer two questions after the distraction phase:  1) ``Recall the 10 words as fast as and as many as possible", and 2) ``Recall the bonus word only". However, they did not know the difficulty of tasks would be controlled. 
To help the participants get familiar with the experimental settings, especially for the four different ways to highlight a bonus word (see Fig. \ref{fig:4conditions}), a trial task was conducted prior to the main experiment. The trial task was based on 1 out of the 8 different cases from the main task. 

In the main experiment, participants were given the eight tasks in random order to avoid carry-over effect. 
During the preparation phase, the participants were asked to focus on the screen for 10 seconds. Then, they were asked to turn around towards to the multi-robot interface for starting the encoding phase.
 In the encoding phase, ten words were presented on the whiteboard of the multi-robot interface. Among the words, one word was defined as a bonus word by the four different ways. The participants were asked to memorize the word set for 60 seconds, and a timer was presented for the participants to keep a track of the time. After 60 seconds, they were asked to turn around towards the screen again, and the distraction phase started.
 In the distraction phase, the participants were asked to verbally answer a series of arithmetic questions which were presented on the screen for 60 seconds. 
 In the retrieval phase, the experimenter asked the two questions  to the participants : 1)``Can you recall the words as fast as and as many as possible?'', and 2)``What was the bonus word?''. Time limit (60 seconds) was given to the participants to answer each question. 

Once all the eight tasks were completed, participants were asked to complete a post-questionnaire about their preference over the four different deictic cues to emphasize the bonus word. 
Based on the answers of the participants on the questionnaire, we conducted the semi-structured interview with a lead-off question to understand their overall experience with the multi-robot interface (e.g., ``Could you briefly share your experience with the entire experiment?'') and follow-up questions (e.g., ``What made it easy or difficult to complete the tasks'', ``Do you think that the four deictic cues affect your task performance?'', ``Did you use any particular strategies to memorize the words?'').
The entire experiment took around 45 minutes, and the participants were individually tested. All sessions were audio/video recorded for further data analysis to track the participants' performance to recall the words and analyzing interview. 
 
\section{Measurements}
To investigate the effect of each condition on memorizing word sets, we collected both quantitative and qualitative data. With the behavioral data, we analyzed participant’s task completion time and the number of correct responses regarding 10 words and a bonus word to compare their performance in each condition. For understanding the participant's subjective experience in the different deictic cues, we analyzed the survey and interview results. Due to the technical errors, we were not able to include data of Participant 5 [P5] and only analyzed 39 participants' behavioral and subjective response. In Section \ref{measure recall word set} and \ref{measure recall bonus word}, we describe the measurements that we used to analyze overall and particular performance of participants, respectively. In Section \ref{measure subjective}, we explain how we examined the subjective preference of the participants over different strategies by analyzing survey and interview data.


\subsection{Performance on recalling a word set} \label{measure recall word set}
For evaluating the general performance, we used two measurements that have been used in traditional memory tasks: an accuracy (the number of correct answers) and time that it took for the subject to make the response (called response time  (milliseconds)) \cite{kahana1999response}. With those two units of analysis, we compared the overall performance of participants across the conditions to recall 10 words as fast as and as many as possible: 1) throughput (unit: 1/msec) and 2) the first recall latency. 

First, to compare the overall performance of each condition, we used throughput units (= the number of correct answers / task completion time) that could compromise speed-accuracy trade-off \cite{drake1996cognitive, thorne2006throughput}. 
Second, to look into how different conditions affect the participants’ search process, we selected the first recall latency on participants’ successive recalls in each condition which is defined by amount of time to recall the first word from the start \cite{rohrer1994analysis}. A long pause before participants verbally recall words means that the participant took time to find the search set/process, and it is a prominent feature of the free recall \cite{rohrer1994analysis, unsworth2008exploring}.

\subsection{Performance on recalling a bonus word} \label{measure recall bonus word}
We applied two different approaches to analyze how different conditions particularly affect the participants’ recall on the bonus word. First, we examined whether the participant recalled the bonus word or not, the accuracy, which is either a 0 or 1. 
Second, we considered the order of recalling a bonus word when the participants retrieved the 10 words, instead of examining the response time to recall the bonus word. This is because participants tended to verbally recall the bonus word immediately after the experimenter asked them. Therefore, it was hard to record the response time due to technical limitations. 
Based on the previous studies that have shown recall order tends to align with the encoding order (or original order of presentation) \cite{kintsch1970models,mandler1969input}, we particularly examined whether the different deictic cues had an influence on recall order of the bonus word. 
In this regard, 
we defined the recall order of bonus words by dividing the recall order of a bonus word into the total number of recalled words. For instance, if the bonus word was recalled at the second order and the participant recalled total eight words, the recall order of a bonus word was 0.25 (= 2/8).

\subsection{Participants' subjective experience} \label{measure subjective}
With the results of the post-questionnaire, we conducted a non-parametric test to investigate the participants' general preference on the four different deictic cues. The survey results presented the participants' preference on the conditions but did not provide an in-depth understanding of the reasons why participants performed better in particular tasks beyond the statistical findings. To investigate the reasons behind their preference and to understand why the participants performed better in specific conditions, we performed a thematic analysis \cite{braun2012thematic} to analyze the interview data. Specifically, we coded the interview data based on the reasons why they felt that a particular task was difficult or easy to complete and how they perceived the multi-robot. 

\section{Results and analysis}
In this section, we will describe statistical findings and present interview results. As we were not able to include data of P5 due to the technical errors, we analyzed 39 participants' behavioral and subjective responses. In a preliminary phase of data analysis, we inquired whether the data followed a normal distribution. When data did not satisfy the normality assumption, we transformed the data to get normality or conducted a non-parametric test. All data were rounded to two decimal places except the throughput which were rounded to five decimal places. 
All statistical analysis was performed with IBM SPSS Statistics 26. 
With interview data, we will elucidate why the statistical outcomes were driven in a qualitative way. To do so, we present several quotes from the participants to support our findings. 

\subsection{Performance on recalling a word set} 
\begin{figure}
    \centering
    \begin{subfigure}[b]{0.8\linewidth}
        \centering 
        \includegraphics[width=1\linewidth]{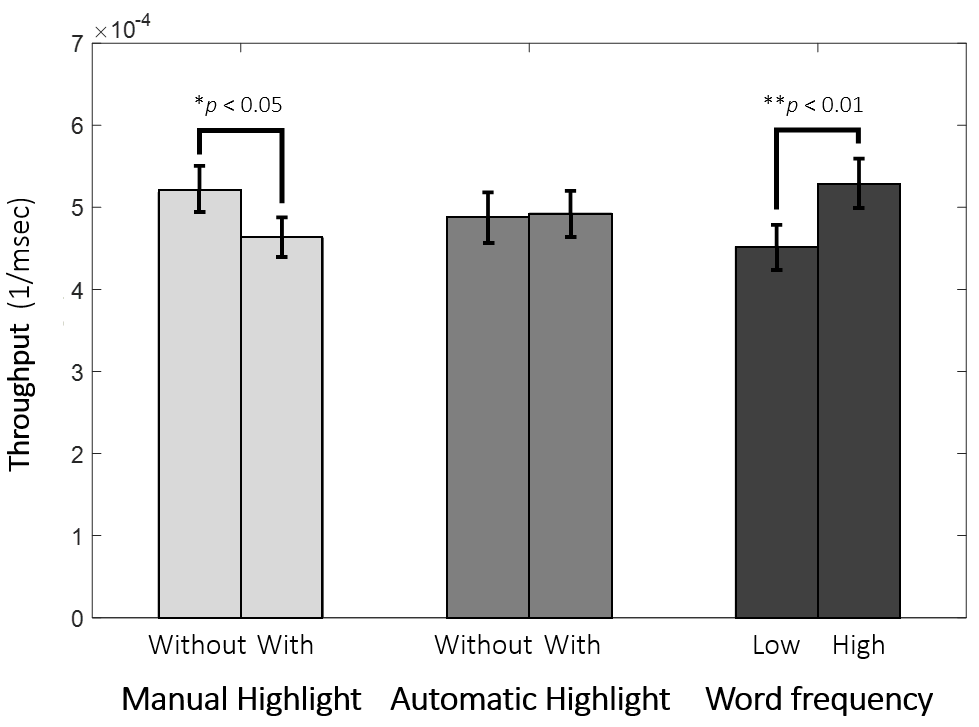} 
        \caption{Throughput}
        \label{}
    \end{subfigure}
    
    \begin{subfigure}[b]{0.8\linewidth}
        \centering 
        \includegraphics[width=1\linewidth]{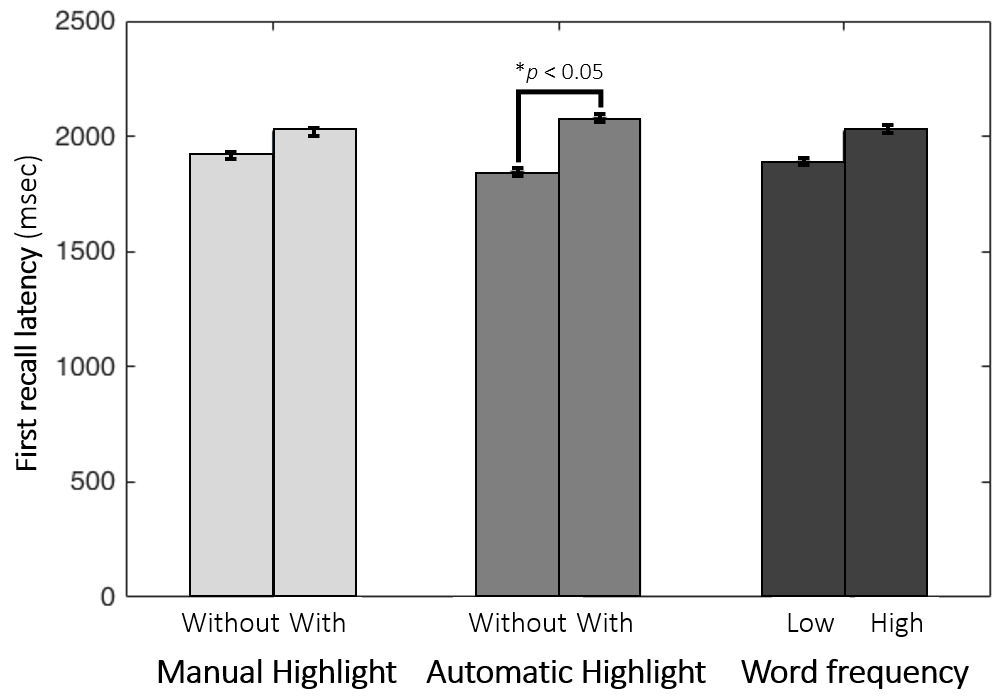} 
        \caption{First recall latency}
        \label{}
    \end{subfigure}
    
    \caption{Comparing throughput and first recall latency across the conditions; error bar represents one standard deviation.} \label{fig:10wordsrecall}
\end{figure}

To compare the participants' performance on recalling a word set across the conditions, we conducted a three-way within-subject analysis of variance (ANOVA). We log10-transformed for the first recall latency since it did not follow a normal distribution. A Greenhouse-Geisser correction was used to adjust for violations of the sphericity assumption \cite{winer1971statistical} and the original degrees of freedom were presented with $\varepsilon$ and $\eta_{p}^{2}$. We used a Bonferroni correction to perform a post-hoc analysis \cite{kirk1974experimental}. In this subsection, we present the results of the throughput and the first recall latency which were examined to the effect of different conditions on the participants' performance on recalling a word set (see Fig. \ref{fig:10wordsrecall}). The displayed values in the graphs are not log-transformed. 

\subsubsection{Throughput}
We found the main effect of manual highlight on the throughput (F(1, 38) = 4.21, p$ < 0.05$, $\eta_{p}^{2}$ = 0.10). The participants performed significantly better when the experimenter pointed out the bonus word with their finger than when the bonus word was manually highlighted by a hand-drawn circle (average of throughput: 0.00052 and 0.00046, respectively).  
Interestingly, however, there was no significant difference in the participants' performance when the bonus word was automatically highlighted by a multi-robot, but the throughput was slightly higher when the multi-robot was used to emphasize the bonus word than when it was not used. 
For the word frequency, as expected, participants outperformed when word sets of the high-frequency words were presented than when those of the low-frequency words were presented (F(1, 38) = 9.99, p-value$ < 0.01$, $\eta_{p}^{2}$ = 0.21). 

\subsubsection{First recall latency}
Participants tended to spend more time to recall the first word when the bonus word was visually highlighted by a multi-robot (without a multi-robot: 1862.09 msec, with a multi-robot: 2041.74 msec) and a hand-drawn circle (without a hand-drawn circle: 1819.70 msec, with a hand-drawn circle: 2089.30 msec) and when the low-frequency words were presented (low-frequency: 1905.46 msec, high-frequency: 2041.74 msec) (see Fig. \ref{fig:10wordsrecall} (b)). However, we only found a main effect of (automatic highlight by) the multi-robot on the first recall latency (F(1, 38) = 4.88, p-value$< 0.05$, $\eta_{p}^{2}$=0.11). It took longer time to recall the first word when the bonus word was highlighted by the multi-robot than the control condition in which the bonus word was only noted by the experimenter. 

\subsection{Performance on recalling a bonus word}
We employed two different approaches to examine the participants' performance on recalling a bonus word. For the recall rate of a bonus word, we conducted a Friedman test \cite{friedman1937use} as the data did not satisfy the normality. For the recall order of a bonus word, we only considered 32 participants' performance who successfully recalled the bonus word in all sessions and performed a repetitive ANOVA. 
\subsubsection{Recall rate of a bonus word}
When the bonus word was highlighted by the hand-drawn circle, the recall rate of the bonus word was higher than any other conditions (average of the recall rate: 1.00 (low-frequent word sets), 0.97 (high-frequent word sets)) (see Table ~\ref{accuracy_bonusword}). However, there were no prominent differences between the conditions (chi-square value = 3.74, degree of freedom = 7, p-value = 0.81). 

\begin{table}
\centering
\caption{Recall rate of a bonus word.}
\resizebox{0.5\textwidth}{!}{
\begin{tabular}{llcc} \toprule
 \multicolumn{2}{l}{Within-subject factors} &{Low-frequency words}  & {High-frequency words}  \\ \cmidrule{3-4}
A hand-drawn circle & A multi-robot & Avg (std)\textsuperscript{a}  & Avg (std)\textsuperscript{a} \\ \midrule
 \multirow{2}{*}{Without} & Without & 0.92 (0.27)  & 0.95 (0.22) \\
 & With & 0.95 (0.22)  & 0.92 (0.27) \\ \\
 \multirow{2}{*}{With} & Without & 1.00 (0.00) & 0.97 (0.16) \\
 & With & 0.92 (0.27) &  0.95 (0.22) \\\bottomrule
\end{tabular}
}

\hfill \textsuperscript{a} Average and standard deviation of recall rate on a bonus word.\\
\label{accuracy_bonusword}
\end{table}

\subsubsection{Recall order of a bonus word} \label{results recall order}
\begin{figure}
  \centering
  \includegraphics[width=0.9\linewidth]{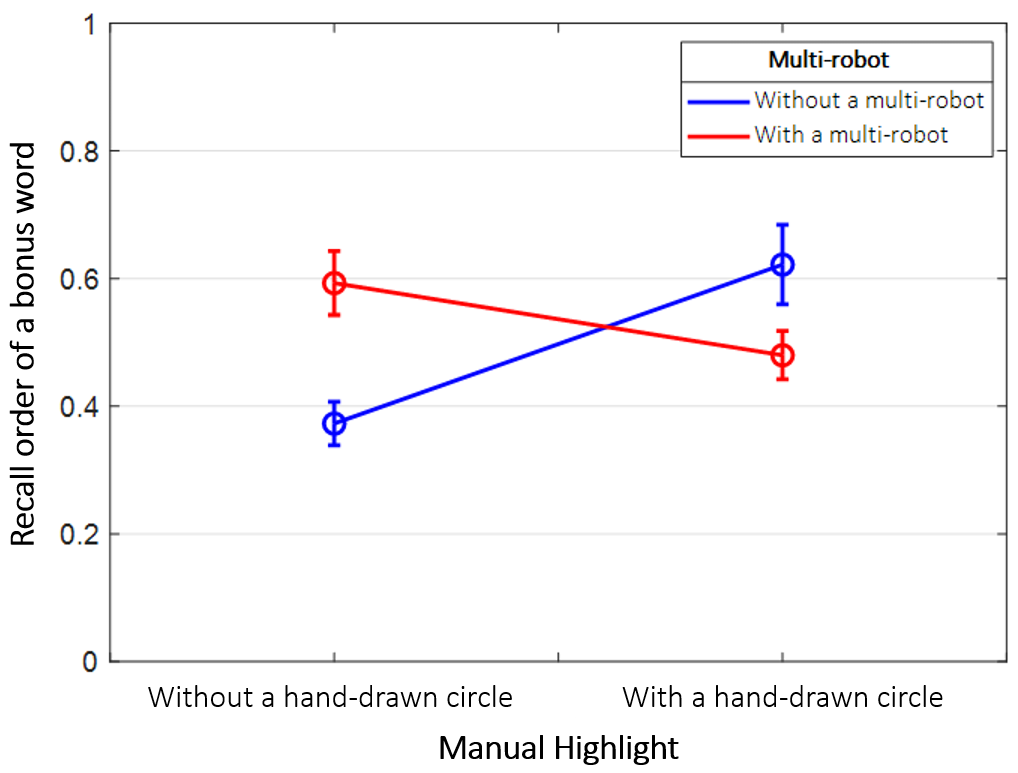}
  \caption{Comparing recall order of a bonus word within two within-subject factors: manual highlight versus automatic highlight by a multi-robot; error bar represents one standard deviation.} 
  \label{fig:recallorder}
\end{figure}

As to the recall order of a bonus word, we found a significant interaction between two within-subjects factors that were applied to present different ways of highlighting the bonus word: a manual highlight and an automatic highlight 
(F(1, 31) = 9.98, p-value $< .01$,  $\eta_{p}^{2}$ = 0.24) (see Fig. \ref{fig:recallorder}). When the bonus word was not highlighted by a hand-drawn circle, the participants recalled the bonus word significantly later when there was a multi-robot than they did in the conditions without the multi-robot 
(condition A: avg = 0.37, std = 0.03, condition B: avg = 0.59, std = 0.05, p-value $<$ .01). On the other hand, when a bonus word was highlighted by a hand-drawn circle, the participants recalled the bonus word slightly earlier if the multi-robot also surrounded the bonus word, but we could not find a significant difference between these two conditions (condition C: avg = 0.62, std = 0.06, condition D: avg = 0.48, std = 0.04, p-value = 0.06).  

\subsection{Participants' subjective experience} \label{results subjective}
In this subsection, we will present the results of the post-questionnaire and elucidate the statistical findings with interview data. Additionally, we will share several notable quotes that show how the multi-robot affects the participants' performance to memorize a word set and a bonus word. Before describing these topics, we want to share the strategies that participants employed to memorize words. Although we did not ask them to apply strategies to perform tasks better, they naturally employed several distinct strategies to memorize the words. They mostly tended to categorize words based on their common features which is well-known strategies to enhance memory \cite{yang2013aging,dunlosky2013improving}. 88\% of participants tried to find semantic association between words. They tried to make stories with the words. 40\% of participants tried to make phonetic association by checking initial of each words or considering sound of words. 70\% of participants said they tried to recall a word by imagining the position of the word on the board.   

\subsubsection{Preferences on different deictic cues}
With the outcome of questionnaire, we performed a Friedman test to investigate the participants’ preference in the four different conditions (see Fig. \ref{fig:4conditions}). There was a statistically significant difference in their preference depending on the experimental conditions (chi-square value=  7.60, degree of freedom = 3, p $<$ .001). For the pair-wise comparisons, Wilcoxon signed-rank tests were conducted with a Bonferroni correction applied (i.e., a significant level at p$<$0.0125). 
There were no significant differences between conditions C and D (C-D: Z= 2.19, p=0.03, \textit{n.s}), and conditions A and B (A-B: Z= 2.11, p=.04, \textit{n.s}). However, the participants significantly preferred conditions D and C than conditions B and A (p$<$.001). 
Participants preferred a clear and long-lasting cue to recognize the bonus word at the beginning of the tasks regardless of the existence of a multi-robot. For instance, P33 mentioned ``I know already what the bonus word is, so I did like `I know this is the bonus word, let's stuck in my mind' and I moved on.'' When it comes to the condition A, participants felt the condition was too weak to inscribe the bonus word on the memory as it did not last for whole 60 seconds. 

\subsubsection{Participants' perception of a multi-robot}

While we conducted the interview, there were various comments on participants' perception on the multi-robot. Based on the thematic analysis, we classified the comments based on design factors of the multi-robot: sounds, motion paths, and movements. Although the motion paths could be under the characteristics of multi-robot movements, we did not merge the theme to the movements as it showed a different tendency of the participants preferences.\\
\textit{Sound of a multi-robot}. The sound of a multi-robot was one of the main reasons why participants did not prefer the conditions with a multi-robot (i.e., condition B and D). Due to the DC motors of the multi-robot, the noises were generated whenever the multi-robot was moving. The participants were sensitive to the sound as they tried to concentrate on the situation to memorize the words. \\
\textit{Motion paths of a multi-robot}. 
Beyond the noise issues with a multi-robot, participants complained that motion paths of a multi-robot distracted them from focusing on the words on the whiteboard. Specifically, the participants pinpointed that the motion paths of the multi-robot sometimes occluded the words while the robots were moving toward the bonus word (``It was distracting because the robots sometimes covered the word. so I had to wait until the robots moved. that was why I chose condition C is the most helpful [P17].''). Those reasons made them prefer the condition C which presented the bonus word in simple and clear ways. 
Although participants gave harsh comments on the sounds and motion paths of the multi-robot as it distracted them from concentrating on the tasks, there were conflicting comments on the movement of the multi-robot in overall.\\
\textit{Movement of a multi-robot}.
Due to the gradual changes of multi-robot movements, participants felt that they wasted time to memorize the words in the condition B when the bonus word was only presented by the multi-robot. This is because they need more time to figure out what the bonus word is in the condition B as it took time for the multi-robot to reach the bonus word (``They (multi-robot) took time to get there and I want to look at the robots instead of the words.[P27].''). 
It implied the movement of the multi-robot definitely draw their attention to figure out the bonus word but was not helpful for them to memorize other words (``It was bothering my process of memorizing. So they're like approaching to the bonus word. I couldn't like focus on the others until they reach out to the word [P19].''). Participants who tended to check the bonus word at the beginning claimed these tensions more frequently and preferred other conditions such as A, C, and D when the bonus word was presented at the beginning of the tasks by other methods rather than the multi-robot. The side effect of presenting a bonus word with the multi-robot was alleviated when the bonus word was already presented in earlier stages just like in the condition D. 
These findings could explain the results in Section \ref{results recall order} that people recalled the bonus word earlier in the condition A than B and the condition D than C as they perceived the bonus word earlier in those conditions. However, some people who did not try to check the bonus word first felt comfortable with spending time to wait for the multi-robot (``I didn't remember the bonus word first begin with. So it (multi-robot) didn't really make a difference [P36].''). 

While participants felt the multi-robot distracted due to its noise and motion paths, they said that the multi-robot was also helpful for them to memorize the bonus word. In interview with P10, he mentioned that the movement of the multi-robot was distracting but helpful for him to check the bonus word. 
Specifically, they compared the other deictic cues used in the experiments and emphasized that the multi-robot gave more salient affordances to recognize the bonus word (``Highlighting is just a static thing. But when I'm memorizing the words, the robots, they're coming and moving. It makes the thing more lively. So it gives me a sense of more concentration to the word [P40].''). Moreover, several participants mentioned that the dynamic movement of the multi-robot guided them to draw a mental image which could help them to recall the whole picture of the board when they tried recalled the word set (``Motion of the robots is helpful. While I tried to recall, I have an image in my head [P12].''). The redundant cues in the condition D were also enhanced the positive effect of the multi-robot on memorizing the word set (``In condition D, the bonus word was ingrained in with my visual memory. I didn't really (try to) remember the word but I remembered [P38].''). 

\section{Discussion}
We conceive that a multi-robot can present a new way of highlighting information. With this in mind, we summarize important findings and discuss implications that shed lights on the value of the multi-robot as a new type of a display in which a deictic cue can be presented through its dynamic movements. 

\subsection{The movement of a multi-robot}
In this subsection, we further discuss the possible reasons that would lead our statistical findings based on our interview data and emphasize the effect of a deictic cue created by a multi-robot on information retrieval. In this study, we asked the participants to perform a dual-task: memorizing 1) a word set of 10 words, and 2) the bonus word. In this regard, we applied different approaches to investigate the effect of the multi-robot on the overall and partial performance of memorizing words by comparing to other deictic cues (i.e., a pointing gesture and a hand-drawn circle).
The reason why we set up two different ways to measure the task performances was to consider a trade-off effect of highlighting information. When a specific part of information is visually highlighted, it not only draws people's focal attention but also hinders people from having a holistic view in the situation \cite{de2009towards}. 

As the exposure time of the deictic cues for a bonus word was different across the conditions, the participants had benefited from the conditions with a hand-drawn circle to memorize the bonus word as it was presented for the longest time (i.e., 60 seconds). However, we could not find any significant main effects of different deictic cues on the recall rate of the bonus word. This result might be driven because the task difficulty of recalling a bonus word was not high enough. Nonetheless, it implies that the participants could compensate the disadvantage in some ways as the bonus word was presented in a shorter period of time in those two conditions (i.e., highlighting the bonus word with a gesture of the experimenter and with the multi-robot, which are the conditions A and B). Interestingly, the deictic cues in the conditions A and B rather significantly affected the participants' performance of memorizing 10 words. Specifically, in terms of the throughput, the hand-drawn circle impeded their performance when we compared to the control condition in which the bonus word was pointed out by a gesture of the experimenter. However, the multi-robot did not hinder the participants' recall on the word set. 

\begin{figure*}[!t]
  \centering
  \includegraphics[width=1\linewidth]{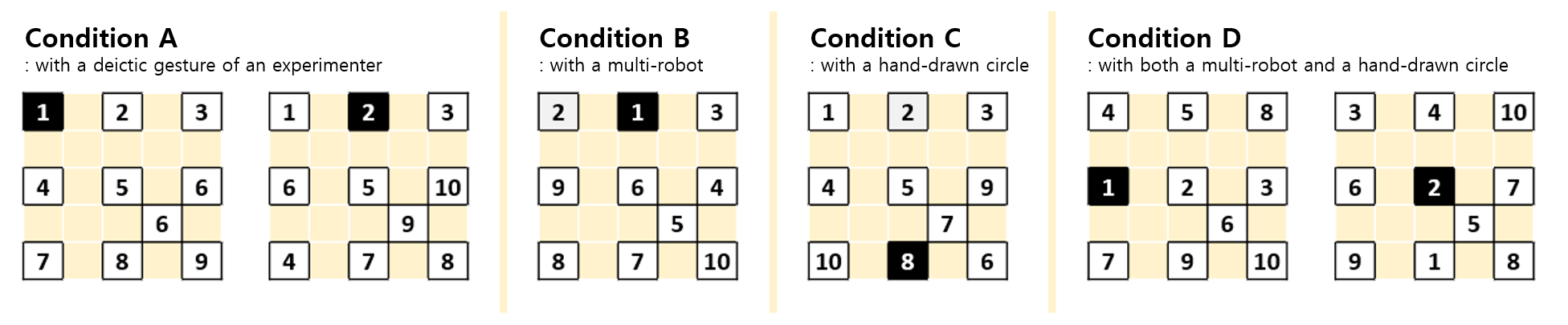}
  \caption{Recall order of each word on the whiteboard; the position of the bonus word is indicated by a black-colored cell, and the recall order of each word is marked with a number in each cell.} 
  \label{fig:board}
\end{figure*}

When we considered that participants spent a significantly longer time to recall the first word in conditions with the multi-robot, this result can be elucidated that the participants recalled more words within a shorter time. In other words, the multi-robot would make them encode the word set as a whole picture. It led to longer latency to recall due to the larger search set \cite{wixted1994analyzing}.
These results may imply that the hand-drawn circle stimulated the participants' selective attention on the bonus word, but it would rather hinder people from making connections between the words by semantic associates \cite{dunlosky2013improving}. According to our interview data, the participants employed to memorize the words was to make a semantic connection between words such as making a story and creating a new word with the first letter of the words which are helpful to recall the words later \cite{nelson1979remembering} (see Section \ref{results subjective}). To make a semantic cluster of words, participants needed to see the whole picture of 10 words. In this regard, the multi-robot might be a helpful medium for this purpose (i.e., memorizing a specific information in a holistic view). That is, the multi-robot may support them to efficiently deploy their distributed attention on the words on a whiteboard \cite{chong2011distributed}. We also found that this interpretation also corresponded with what participants mentioned in the interview. One participant described that he could recall the whole picture of the whiteboard with 10 words when the bonus word was highlighted with the multi-robot: ``Motion of the robots is helpful. While I tried to recall, I have an image in my head [P12].'' 
Although the movement of the multi-robot did not highly impact on the participants' performance to recall the words, we could find that the multi-robot would reshape the way people perceive information and help them use their cognitive resources effectively when they have to complete a dual-task.

\subsection{Encoding and presenting order}

Based on the previous studies that have shown the recall order tends to be aligned with the encoding order 
\cite{kintsch1970models,mandler1969input}, we examined whether different deictic cues had an influence on the recall order of the bonus word in Section \ref{results recall order}. We found that a multi-robot was helpful for the participants to recall the bonus word earlier only if the bonus word was already highlighted by a hand-drawn circle. Through the interview with the participants, we noted that the recall order of words was also affected by how the words were presented on the whiteboard which corresponded with the previous studies  \cite{kintsch1970models,mandler1969input}, because one of the main strategies that participants used to memorize the words was related to the location of the bonus word on the whiteboard. 

In this subsection, we present an additional analysis to investigate how different deictic cues would guide the recall order of the words by considering the position of the words. We visualized the recall order of the 10 words based on their positions on the whiteboard and investigated whether there were particular patterns between the position of the bonus word and the way of recalling the 10 words. We did not consider the factor for the word frequency because there was no significant difference on the recall order of a bonus word. Therefore, we compared the recalling patterns across the four different conditions regarding two factors: 1) manual highlight with a hand-drawn circle, and 2) automatic highlight with a multi-robot (see Fig. \ref{fig:4conditions}). 
To investigate the effect of different deictic cues on the recall patterns, we first classified cases based on the position of the bonus word. 
We focused on the representative cases which occupied more than 25\% of the overall cases. In conditions B and C, we could select one representative case respectively as there were prominent cases composing more than 80\%. However, we chose two cases respectively in conditions A and D due to the similar occurrence in each case. Second, for each scenario, we calculated the average of the recall order of words on each position of the whiteboard by dividing the sum of the recall order of the words on the designated position into the occurrence of successfully recalling the words. Third, we ranked the (10) positions in order of the average. Lastly, we visualized the recall order and investigated the patterns (see Fig. \ref{fig:board}). 

Overall, we found that participants tended to recall the words just similar to how the words were presented on the whiteboard. As shown in Fig. \ref{fig:board}, participants tended to recall the words from top right to bottom left in all conditions except the condition D when the bonus word was surrounded by both a hand-drawn circle and a multi-robot. On the other hand, the participants tended to recall the words that were located near the bonus word in the condition D. This revealed a different pattern of the recall order in the condition D. As we only analyzed the participants' behavioral data for this analysis and the position of the bonus word was different across the conditions, we could not draw a concrete conclusion. Nonetheless, we could find that the prominent highlights created by the multi-robot and the hand-drawn circle led different patterns of the recall order. This would be further explored in future studies to validate the effects of movements of the multi-robot on information encoding.

\section{Conclusion}
As the collective behavior of multi-robots has benefits for the efficient performance of complex tasks, applications involving the multi-robots have mainly focused on industrial and critical contexts such as manufacturing, construction, and search and rescue. Going beyond the functional aspects of such systems, we were particularly interested in a potential application of multi-robots in everyday life where people use a deictic cue to draw attention to specific information in order to express their intentions. To realize this, we developed a multi-robot-based interface that can be used as a new type of visual display in which a deictic cue is presented through the dynamic movements of the multi-robot. In this study, we examined the effect of the multi-robot on information retrieval by applying a basic paradigm from the psychological study of memory. Specifically, we investigated whether a deictic cue created by the multi-robot aided people in information retrieval and conducted a user study consisting of free recall tasks with three within-subjects factors were considered: 1) manual highlight, 2) automatic highlight, and 3) word frequency. First two factors were to create a deictic cue either by a hand-drawn circle or a multi-robot. The last factor was to control task difficulty. 
Although there was no significant difference in participant's retrieval rate of the bonus word between different deictic cues, we found a significant interaction effect between the manual and automatic highlight on recall order of the bonus word. If there was no a hand-drawn circle, the participants recalled the bonus word earlier if the bonus word was presented at the beginning (i.e., condition A) than if the bonus word was presented with a dynamic movement of the multi-robot (i.e., condition B). However, when a deictic cue was already presented with a hand-drawn circle, the existence of the multi-robot did not affect the recall order of a bonus word. 
We also found that participants in the multi-robot condition took significantly longer to recall the first word but this did not cause a significant difference in their overall performance at recalling the word set (i.e., throughput). This means that the multi-robot would be helpful to recall as many as words after the first recall. We also noted standard deictic cues to have side effects that rather hindered participants recall of the work set; however, the deictic cue created using the multi-robot did not affect overall performance. 

While we did not identify a significant effect of the multi-robot cue on information retrieval, the interview data provided several implications that could enhance our results in future studies. Participants commented that the kinesthetic movements of the multi-robot helped them to recognize the bonus word and recall the whole image of the board; that is, the multi-robot created a prominent affordance. However, they also reported the multi-robot irritated them with its sound and occlusion issues (one of the limitations of our study), thus the basic need to focus on the task was not adequately supported. As people felt less distracted when they knew the direction or goal position of the multi-robot, it is also important to consider auxiliary information that could lessen user concern regarding the multi-robot. Moreover, how the characteristics of the multi-robot affect information encoding and retrieval should be further investigated because the movement and formation of the multi-robot when it presents information would affect a user’s cognitive and emotional status \cite{santos2019motions,le2016zooids}. In future studies, we would like to explore how a multi-robot can guide the information encoding process by using eye tracking information. We hope that our study can extend interest in designing a multi-robot system as a platform to present information and guide communication by using dynamic formations to provide prominent cues. 


\bibliography{main}
\bibliographystyle{IEEEtran}

\end{document}